\documentstyle[11pt,aaspp4]{article}
\slugcomment{ApJ, in press}
\lefthead{Andreon}
\righthead{Is there a deficit of S0s at intermediate redshift?}
\begin{document}

\title{Is there a deficit of S0s at intermediate redshift?\footnote{Based
on observations obtained with the NASA/ESA {\it Hubble Space Telescope}}} 
\author{S. Andreon}
\affil{Osservatorio Astronomico di Capodimonte, via Moiariello 16, 80131
Napoli, Italy \\ email: andreon@na.astro.it}

\begin{abstract} 

Two contradictory results on the evolution of SO galaxies now exist in the
recent literature; either S0s are old ($z_{formation}>2$) and are
evoloving passively or most of them formed at $z<0.5$, as implied by the
deficit of S0s in intermediate redshift ($z\sim0.5$) clusters.  The
resolution of this controversy may be that the apparent deficit of S0s has
been derived from a quantity -- the E to S0 ratio -- which is prone to
morphological classification errors. Once all sources of error are taken
into account, the E to S0 ratios of clusters at different redshifts are
fully compatible, and no additional creation of S0s at $z<0.5$ is required
by the data. Furthermore, there is no deficit at all of S0s in the
intermediate redshift cluster for which we have morphological types of
very high quality, and thus derive an E to S0 ratio with a small error.

\end{abstract}


\section{Introduction} 

The S0 class is often considered a class of transition between Es and Ss
(see, for example, Hubble 1936, but see van den Bergh 1976 for an opposite
opinion). Evidence for the evolution of the S0s was cast
looking for an heterogeneity of this class, heterogeneity that 
suggests a variety of evolutionary paths which end in a S0 galaxy. 
Color and color dispersion of S0s (Sandage, \& Visvanathan 1978; Bothun,
\& Gregg 1990; Bower, Lucey, \& Ellis 1992),  segregation in clusters
(Oemler 1974; Dressler 1980b), location in the fundamental plane
(Saglia, Bender, \& Dressler 1993), geometrical properties (Michard
1994), the presence of unusual features such as shells, ripples
(Schweitzer et al., 1990): all these factors give contradictory
indications on the origin of S0s galaxies. 

Recently, the advent of the {\it Space Telescope} has permitted a major
breakthrough: its superb angular resolution allows the morphological
classification of distant galaxies and therefore the comparison of the
properties of the morphological types at quite different redshift, i.e. at
different look-back times. The evolutionary history of the morphological
types should no more be indirectly inferred by the relics it leaves in
the galaxy properties, but can be caught in the act. 

Two recent papers by the MORPHS collaboration (Smail et al. 1997; Dressler
et al.  1997) present evidences for the evolution of the S0s class between
$z\sim0.5$ and $z\sim0$. MORPHS noted that in intermediate redshift
($z\sim0.5$) clusters there is a deficit of S0s with respect to nearby
clusters. 

This situation is a quite puzzling: intermediate redshift lenticulars are
red and with little scatter in color, and therefore are thought to be as
old as ellipticals (Ellis et al. 1996; Andreon, Davoust, \& Heim 1997).
Nevertheless their number seems very low, so that many of them must be
created in order to reach the population of S0 in nearby clusters
(Dressler et al. 1997). Furthermore, at least Coma lenticulars are too old
and homogeneous in their properties to be formed at $z<0.5$ (Bower et al.
1992; Andreon 1996). Therefore, S0s in clusters should have
$z_{formation}<0.5$ and $z_{formation}>2$ at the same time. 

The deficit of S0s is not shared by all clusters. In one of the three
clusters at $z\sim0.3$ studied by Couch et al. (1998) the S0s fraction is
quite similar to that found in nearby clusters, whereas in the two other
it is intermediate between nearby and $z\sim0.5$ clusters. In the cluster
Cl0939+4713 ($z\sim0.4$) S0s are as frequent as in the nearby Coma cluster
(Andreon et al. 1997a), a result which would imply no evolution with time
for the S0 population.  The situation is striking for this clusters since
Dressler et al. (1997) and Andreon et al. (1997a) find opposite results
for the same sample of galaxies and from very similar analysis of the same
WFPC2 post-refurbished images of the {\it Hubble Space Telescope}. 

Therefore, a further analysis is useful in order to understand
these discrepancies. 

\section{A deeper look to the data}

Since the spiral population in clusters evolves with redshift (Dressler et
al. 1994; Oemler, Dressler, \& Butcher 1997; Dressler et al. 1997; Andreon
et al. 1997a), the evolution of S0s has to be inferred from the
variation of the ratio between the E and S0 populations. 

The typical composition of nearby clusters, as computed by Dressler et al.
(1997) for a large sample of galaxies drawn from a catalog with undefined
completeness is quite similar to the composition of the nearby Coma cluster
computed by Andreon et al. (1997a): E:S0:S=25:42:31. The latter
composition has been computed adopting the same absolute magnitude limit
and the same rest--frame passband used for intermediate redshift galaxies. 

Intermediate redshift clusters have a typical composition of
E:S0:S=37:19:44 for the $\sim600$ galaxy sample of Dressler et al. (1997)
whereas Andreon et al. (1997b) find E:S0:S=15:34:43 for a smaller sample
of $\sim70$ galaxies in Cl0939+4713.  The two E to S0 population ratios at
$z\sim0.4$ computed from these cluster compositions are statistically
different at the 99.999 \% confidence level.  Therefore, the two E to S0
ratios do not differ because of small number statistics. Dressler et al.
(1997) estimate the E to S0 ratios of intermediate redshift clusters and
of nearby clusters as statistically different, while Andreon et al. (1997)
find them equal.  Therefore the disagreement between the two works and the
puzzle of the deficit of S0 galaxies in intermediate redshift clusters is
not due to differences in the zero-redshift samples or to small number
statistics. We need now to analyse the possible existence of other sources
of errors in the E to S0 ratio. 

A possible one comes from morphological misclassifications.  The error
(scatter) on the MORPHS morphological types of intermediate redshift
galaxies is quantified by Smail et al. (1997) in their Figure 1. About
20\% of all galaxies in the magnitude range of interest (brighter than the
classification limit adopted in estimating the E to S0 ratio), have
morphological estimates which differ, from morphologist to morphologist,
by more than one class in the revised Hubble scheme, i.e. the difference
between an E and a S0 or between a S0 and a Sa. The error on the
morphological types claimed by MORPHS is not small enough to exclude that
the S0 population is the same at $z\sim0.5$ and in the local universe. In
fact, an error of one morphological bin for only 20 \% of the galaxies
(for instance moving 10 \% of Es and 10 \% of Ss at $z\sim0.5$ in the S0
class) would give the zero-redshift ratio. Even smaller errors on the
morphological classification are sufficient to explain the values and the
scatter of the E to S0 ratios found by Couch et al. (1998) in three
clusters at $z\sim0.31$. 

If, at the difference of MORPHS, we take into account the error on the
morphological type, which is the dominant factor, in the computation of
the error on the E to S0 ratio, then the MORPHS E to S0 ratios at
different redshifts are equal within the errors. Therefore, we can
conservatively state that there is, up to now, no evidences for a deficit
of S0s in intermediate redshift clusters.  The evolution of the cluster S0
population is not required by MORPHS data (or better, by the present
analysis of the data). 

Systematic errors are even more dangerous. Here, as well explained by
MORPHS, we are interested to differential morphological errors, i.e. to
possible factors which render the morphological classification dependent
on redshift. Classification errors equally affecting nearby and
intermediate redshift clusters (e.g. the misclassification of face-on
lenticulars as ellipticals, Capaccioli 1987) do not affect the ratio of E
and S0 populations, since they cancel out in the comparison, as long as
the fraction of misclassified galaxies is the same at all redshifts. As
shown in Andreon et al. (1997a) and Dressler et al. (1997), images of
intermediate redshift clusters from the space resemble very closely to
ground based images of nearby clusters in terms of rest-frame deepness,
resolution and sampled passband. This seems to exclude an important
differential effect on the determination of the morphological composition
due to the different nature of the images. However, given the subjective
nature of the morphological classification this is not fully guaranteed. 
 
In order to investigate the possible existence of systematic errors in the
morphological classification, we compare MORPHS morphological scheme, based on
the resemblance of galaxies to standards, with a classification method
which relies on the detection of structural components (such as disk, bar,
spiral arm, bulge, halo, etc.). Two clusters are availables for
this exercise: the nearby Coma cluster (whose Hubble types are listed in
Andreon et al. 1996, Andreon, Davoust, \& Poulain 1997) and the
intermediate redshift Cl0939+4713 cluster (whose Hubble types are listed
in Andreon et al. 1997a). 

At this stage one has to comment upon the definition of the E and S0
types. According to de Vaucouleurs (1959) and Sandage (1961) the
segregation between E and S0 types is based upon the examination of the
surface brightness profiles along the major axis of the galaxy. The
presence of a disk gives a characteristic bump above the radial profile
typical of pure spheroidal galaxies. We classify a galaxy as S0 if it has
not spiral arms or irregular isophotes and if it presents a bump in its
major axis surface brightness profile, which is absent or at most hinted
in the minor axis profile (see for example DG250, DG303 or DG 310 in
Figure 1), in the spirit of the Hubble definition for this class (see
Andreon, \& Davoust 1997 for details). MORPHS discriminate between E and
S0 according to their resemblance to their respective morphological
standards. 

The morphological type of Coma galaxies provided by Dressler (1980 and
used by MORPHS) and Andreon et al. (1996, 1997b) agree well, within the
usual 20 \% of scatter (Andreon, \& Davoust 1997) and the E to S0 ratios
computed from these data are very similar, as expected given the agreement
between these authors on the morphological composition at the
zero-redshift. For certain or likely members of the Cl0939+4713 cluster,
the structural and traditional morphological types show a similar scatter:
23 \% of galaxies have different types (15 galaxies out 65), when they are
binned in the usual three classes (E, S0 and S), while peculiar galaxies
are neglected. The disagreement drop to 18 \% if we do not count as truly
discrepant galaxies whose MORPHS type is uncertain between the structural
type and another type. There is therefore the same agreement between
structural and traditional types (at most 23 \% of them differs) as among
the types estimated by traditional morphologists (20 \% of them differs),
thus showing that structural types are not worse than traditional ones. 

However, Cl0939+4713 is S0--rich using structural types, whereas is
S0--poor using traditional types. Since the existence of a deficit of S0s
in Cl0939+4713 cluster strongly depends on which classification scheme is
used and both schemes are equally good, we can safely conclude that the
claimed deficit of S0s in this intermediate redshift cluster has been
derived from a quantity, the E to S0 ratio, which is prone to
classification errors. 

Let us now move to a detailed comparison of galaxies with discrepant types.
Out of 15 galaxies with discrepant structural and traditional
morphological types in Cl0939+4713, 13 are classified as S0s in Andreon et
al. (1997a) and as Es (7) or Ss (6) by MORPHS. 

Figure 1 shows the major and minor axis surface brightness profile of the
7 galaxies in Cl0939+4713 whose structural type is S0s but are classified
as Es by MORPHS. The curvature of the major axis surface brightness
profile and the linearity of the minor axis profile of 3 of them (DG250,
DG303, DG310) are unquestionable. These features are less evident, but
still present, in DG314, DG443, DG457. In DG302 the bump can be hinted at
$r^{1/4}\sim0.75$ arcsec. S0s in Coma and in Cl0939+4713, as classified by
means of their structural type, show similar ellipticity profiles and
$cos(4\theta)$ deviations of the isophotes from perfect elliptical shape
(see Andreon et al. 1997a for details). Therefore these galaxies are much
more likely to be lenticulars rather than ellipticals. Similar
classification problems are also mentioned by two of MORPHS (Couch et al.
1998). 

Figure 2 shows the images of the six galaxies whose structural type is S0,
thought their traditional type is S. The MORPHS types for these galaxies
are S0/a, Sb and Sbc (2 cases each).  Any spiral arm or irregularity,
which characterize them as Ss, is visible in their images. If spiral arms
were present we would see them. In fact, Figure 3 shows that all other
early type spirals, i.e. Sa and Sbc (there are no Sab or Sb in the studied
sample), as classified by MORPHS in Cl0939+4713, shows spiral arms, thus
showing that the signature we are looking for is visible in the available
images. The images that we compare contain galaxies of similar magnitudes.
They are also part of the same image of Cl0939+4713 and they are shown
with the same cuts, and they are therefore fully comparables. To
summarize, these six galaxies with discrepant types in the two works are
correctly not classified as spiral by Andreon et al. (1997a), since spiral
arms, when present, are visible in the available image. 

In conclusion, in individual galaxies for which structural and traditional
types differ, the structural type gives a better description of the galaxy
appearance, avoiding the classification of galaxies without spiral arms as
Ss and of galaxies with a bulge and a disk as Es.  To give a larger weight
to these types, as justified in previous paragraphs, means to find no
deficit at all for S0s in the intermediate redshift cluster Cl0939+4713.
For this clusters, any S0s should be ``created" to reach the S0s
population of nearby clusters.

\section{Discussion and Conclusions}

Smail et al. (1997) and Dressler et al. (1997) are aware that the claimed
deficit of S0s in intermediate redshift clusters holds only in the absence
of redshift dependent errors in their morphological classification.  For
quantifying systematic errors, they compared the ellipticity distribution
of the morphological types at intermediate redshift and of the nearby
universe.  As nearby sample they used the Revised Shapley-Ames catalogue
(Sandage, Freeman, \& Stokes 1970) and a magnitude complete sample of
galaxies in Coma (Andreon et al. 1996), the latter being preferable (Smail
et al. 1997) because the ellipticities were measured approximatively at
the same isophote of the compared sample. MORPHS finds that the
ellipticity distribution of intermediate redshift and nearby Hubble
classes are compatible, and conclude that, in spite of the 20 \% of
scatter in the morphological types present in their sample, their
classification is equally good and with no bias with redshift. However,
this comparison is intended to look for redshift dependent trends in the
morphological classification by means of an indirect quantity, the
ellipticity distribution of the Hubble types, and does not test directly
the quantity of interest, the redshift dependence of the morphological
classification, as we do in our comparisons for Coma and Cl0939+4713
morphological types. 

To summarize, we have shown that the E to S0 ratios of nearby and
intermediate redshift clusters are equal within the errors once the error
on the morphological type, which is the dominant term, is taken into
account. The comparison of equally good morphological types measured by
means of independent morphological schemes of the galaxies in CL0939+4713
confirms for this cluster that the claimed deficit of S0s is the result of
having measured it by a quantity too prone to morphological errors.
Therefore the evolution of the S0 fraction from $z\sim0.5$ to the present
time is not longer required by the MORPHS data (or better by the present
analysis of the data), thus solving the puzzle of the old age of the S0s
and their absence at $z\sim0.5$. Furthermore, there is no deficit at all
of S0s in the intermediate redshift cluster for which we dispose of high
quality morphological types and thus of a E to S0 ratio with small error. 

However, our conclusion should not be overinterpreted. First of all, we
have not shown the absence of an evolution in all the properties of early
type galaxies, from $z\sim0.5$ to the present time. We have just shown
that there is no evidence for an evolution of the relative fraction of
early type galaxies. Strickly speaking, the constancy of the E to S0 ratio
does not exclude morphological changes of individual galaxies between
these two classes, and from S to E or S0 classes, but just gives some
constraints on the evolution of the two populations. Individual galaxies
can change their morphological type while keeping the E to S0 ratio
constant, provided that the same number of Es become S0s as viceversa, or,
when spiral galaxies are involved, that S become E or S0 with the right
frequency. Furthermore, a 20\% of scatter (error) in the morphological
type, claimed both by MORPHS and by ourselves (this work and Andreon, \&
Davoust 1997) is not a negative judgement of the MORPHS work, since it is
usual in all morphological studies -- so usual that better agreements are
suspicious (Andreon, \& Davoust 1997)--. The presence of systematic errors
in the morphological classifications does not make MORPHS types useless,
since they are still useful for studying quantities less affected by
systematic errors. A better morphological type for all intermediate
galaxies, such as a structural one, would need a 60 time larger effort
(Andreon \& Davoust 1997) and thus is out of our present capabilities. 

Recently, Van Dokkum et al. (1998) found that at large clustercentric
radii ($R>0.7 h^{-1}_{50}$ Mpc) of the intermediate redshift cluster
CL1358+62, S0s are heterogeneous in color and therefore experienced star
formation until very recently, reaching conclusions opposite to those from
previous works focussed on the central regions of other intermediate
redshift clusters (Ellis et al. 1996; Andreon, Davoust, \& Heim 1997).
They suggest that S0s evolve primally in the transition region between the
cluster and the field, giving support to MORPHS finding that S0s are still
forming at intermediate redshift. However, van Dokkum et al. (1998) choose
to classify as S only starforming galaxies, leaving spirals with faint
smooth spiral structure in the S0 class (see their Section 2.3.1). This is
confirmed by the inspection of their black and white prints presented in
their Figure 3: at least 25 \% of all galaxies that van Dokkum et al.
classify as S0s are S according to MORPHS or our morphological scheme. All
these galaxies present smooth spiral arms (or irregular isophotes) and
look as the spirals in CL0939+4713 shown in Figure 3 or to Coma spirals. 
Therefore, Van Dokkum et al. (1998) adopt a classification scheme
different from the one adopted by other researchers.  The shape of the
azimuthal averaged surface brightness profile measured by van Dokkum et
al. (1998) does not discriminate S0s from early--type spirals, since the
difference between the two types is given by the smooth spiral arms whose
contribution to the radial surface brightness profile is negligible. Their
conclusion about the evolutive nature of S0s is therefore relative to
their S0 class, and not to the Hubble S0 class. A reclassification of van
Dokkum et al. S0s in the Hubble scheme would help to understand which part
of the heterogeneity of their S0s class is due to the pollution by
early--type spirals and which part is intrinsic and suggestive of recent
formation. At small clustercentric radii, where the misclassification is
likely low due to the morphological segregation, van Dokkum et al. S0s are
an homogeneous old population of galaxies, as in all the other studied
clusters both in the nearby universe and at intermediate redshift. 

As a final remark, we stress that the morphological type is a quantity
which needs to be calibrated and with an associate error, as all other
physical quantities. We re-iterate the need of using a classification
method which keep minimal morphological type errors and systematic
differences from the standard scale: the Hubble sequence. 

\acknowledgements

I warmly thank E. Salvador-Sol\'e for useful suggestions, L. Buson and
M. Capaccioli, S. Zaggia and, in particular, G. Longo for their comments
to the manuscript that greatly helped to improve the presentation of the
results. I thank Dr. G. Bothun for his adviced suggestions. A particular
acknowledgement goes to the MORPHS team, without who it is likely that we
would not dispose of the images of intermediate redshift clusters whose
observations are so fine tuned for a correct comparison with nearby
clusters.

\newpage

\figcaption[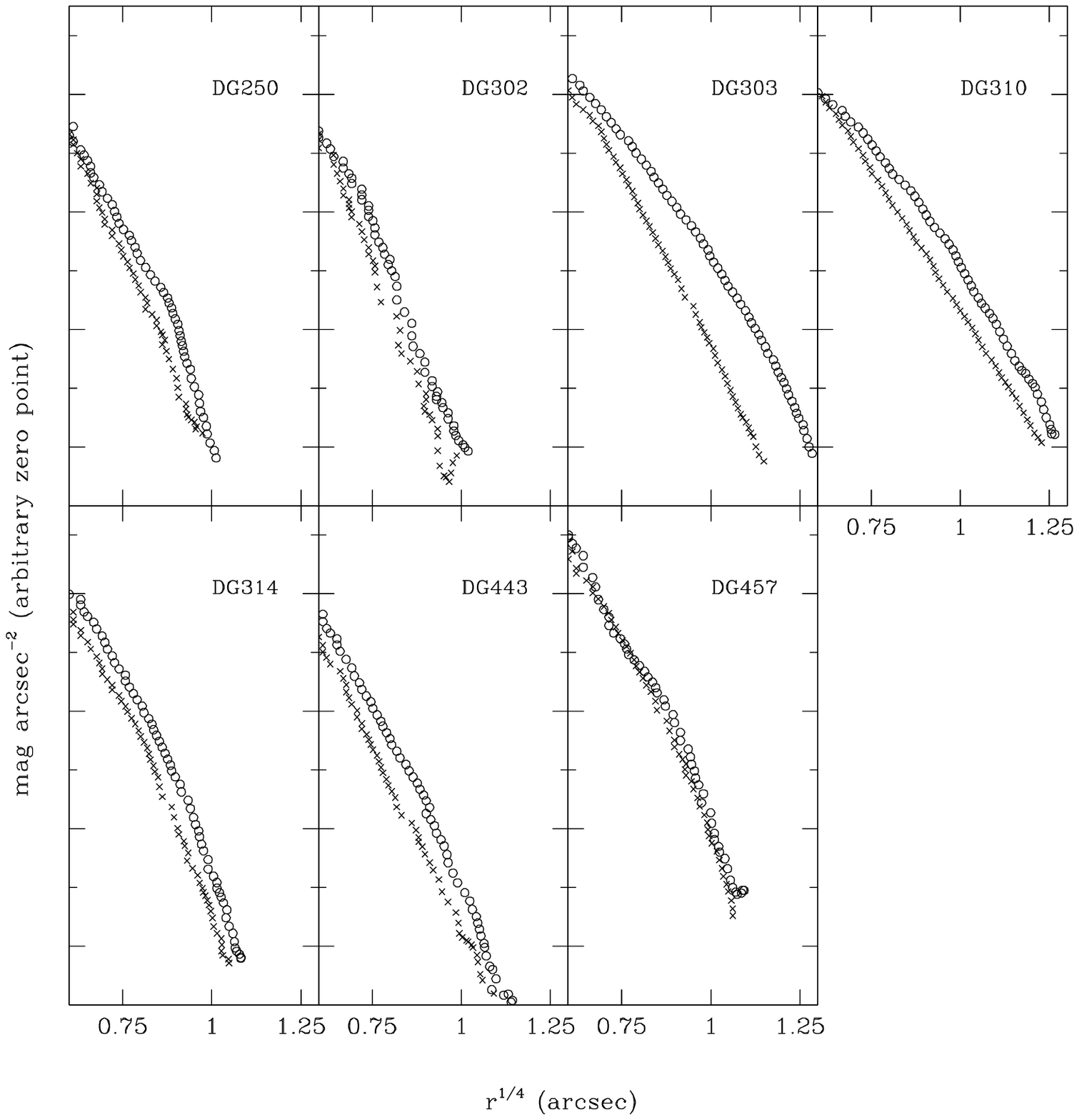]
{Major (circles) and minor (plusses) axis surface brightness profiles for
galaxies whose structural type is S0, but traditional type is E.}

\figcaption[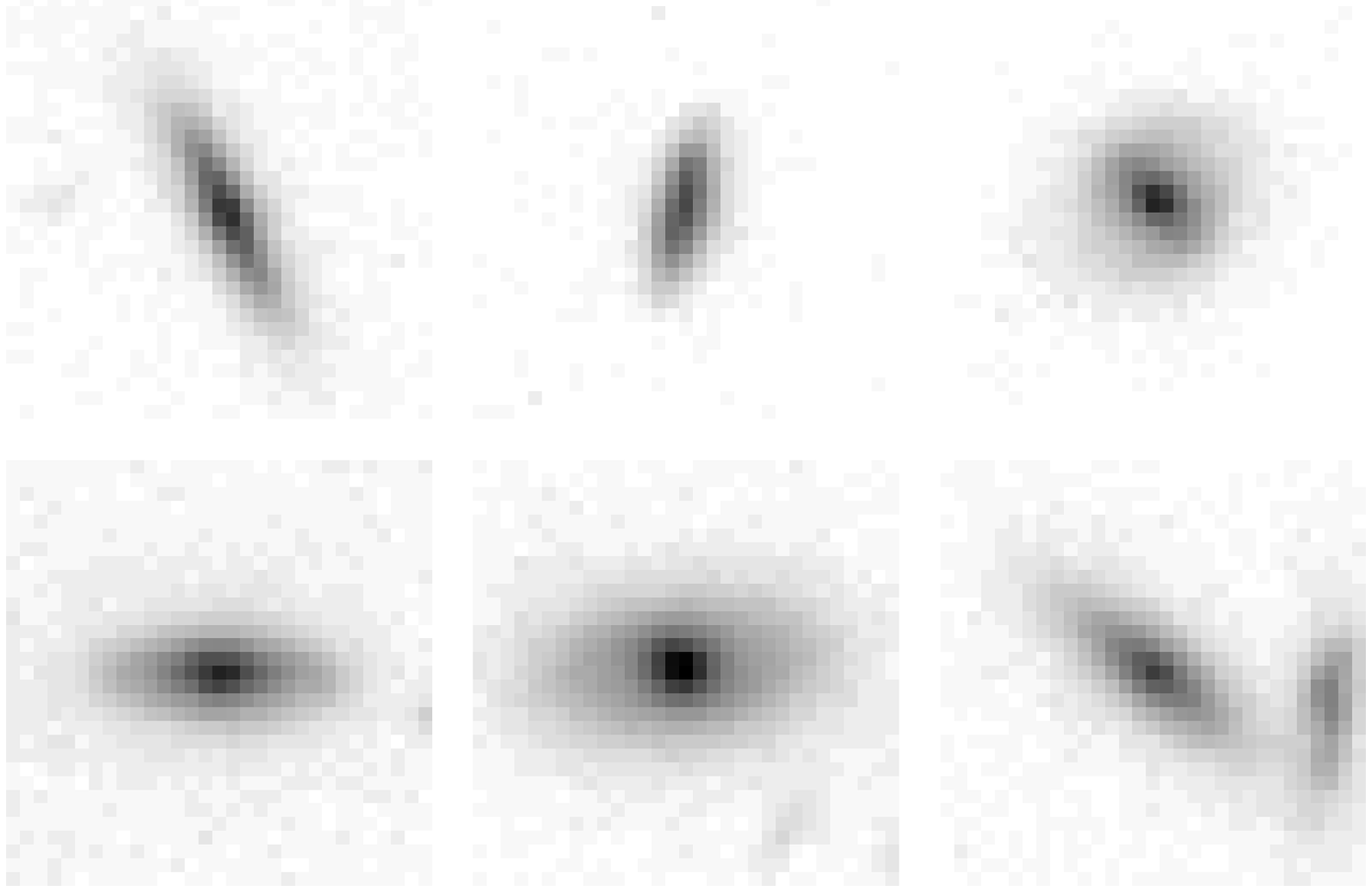]
{Mosaic of the F702W images of the galaxies whose structural type is S0,
and traditional type is S.}

\figcaption[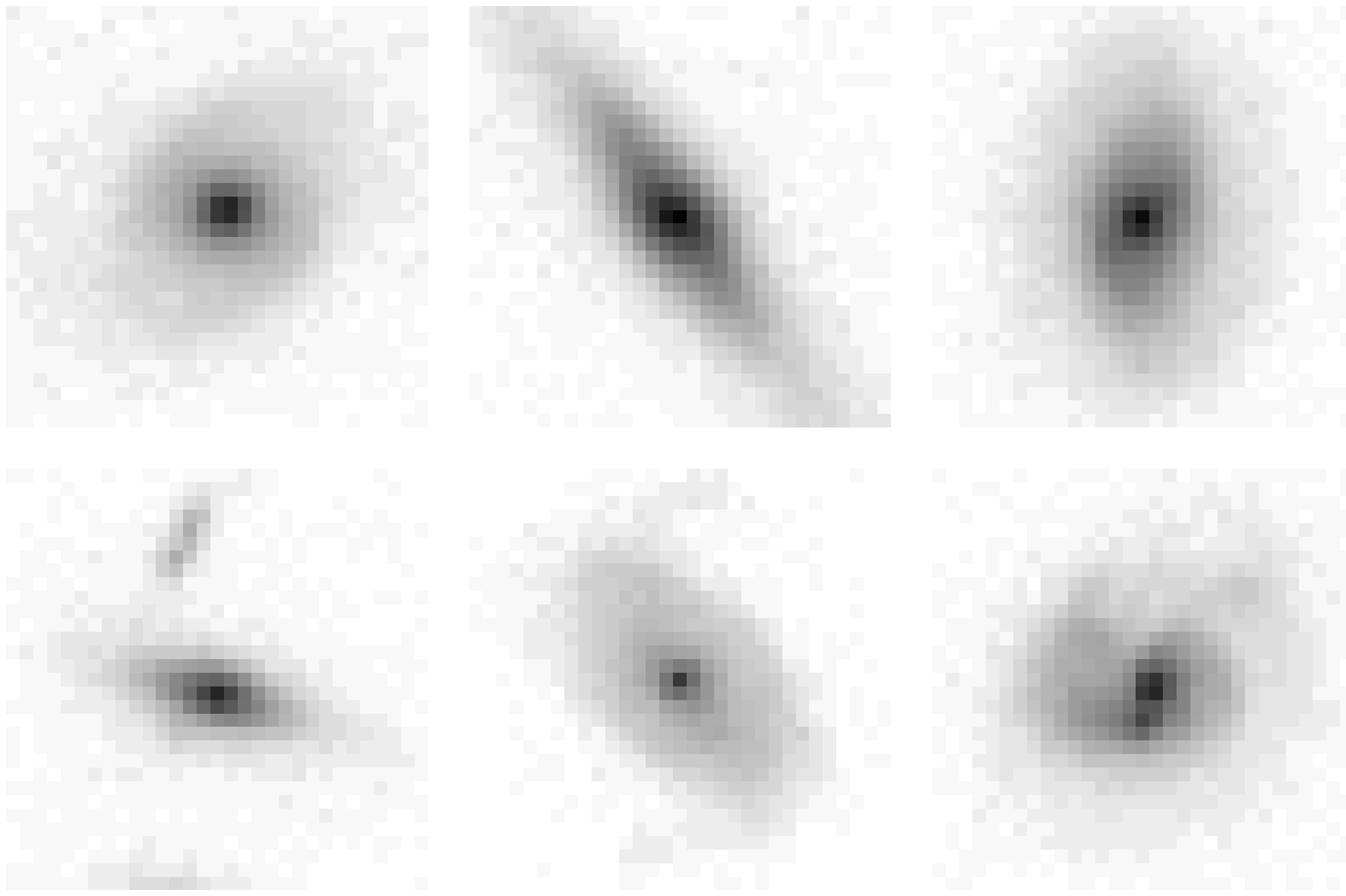]
{Mosaic of the F702W images of the galaxies whose structural and traditional
type is S. Images are shown with the same cuts and look up table of the
previous figure and have been extracted from the same image.} 


\begin{thebibliography}{}

\bibitem[]{}
Andreon S., 1996, A\&A 314, 763

\bibitem[]{}
Andreon S., Davoust E., 1997, A\&A 319, 747

\bibitem[]{}
Andreon S., Davoust E., Heim T., 1997a, A\&A 323, 337

\bibitem[]{}
Andreon S., Davoust E., Poulain P., 1997b, A\&AS, 125, 1

\bibitem[]{}
Andreon S., Davoust E., Michard R. et al., 1996, A\&AS, 116, 429

\bibitem[]{}
Bothun G., Gregg M., 1990, ApJ 350, 73

\bibitem[]{}
Bower R., Lucey J., Ellis R., 1992, MNRAS 254, 601

\bibitem[]{}
Capaccioli M., 1987, in IAU Symposium 127, ed. T. de Zeeuw (Dordrecht;
Reidel), p. 47

\bibitem[]{}
Couch W., Barger A., Smail I., Ellis R., Sharples R., 1998, ApJ, in press
(astro-ph/9711019) 

\bibitem[]{}
Dressler A., 1980a, ApJS 42, 565

\bibitem[]{}
Dressler A., 1980b, ApJ 236, 351

\bibitem[]{}
Dressler A., Oemler A., Couch W., et al., 1997, ApJ, 490, 577

\bibitem[]{}
Dressler A., Oemler A., Sparks W., Lucas R., 1994, ApJ 435, L23

\bibitem[]{}
Ellis R., Smail I., Dressler A. et al., 1997, ApJ 483, 582

\bibitem[]{}
Hubble E., 1936, The Realm of the Nebulae, (New Haven: Yale
University Press)

\bibitem[]{}
Michard R., 1994, A\&A 288, 401

\bibitem[]{}
Oemler A., 1974, ApJ 194, 1

\bibitem[]{}
Oemler A., Dressler A. Butcher H., 1997, ApJ 474, 561 

\bibitem[]{}
Saglia R., Bender R., Dressler A., 1993, A\&A 279, 75

\bibitem[]{}
Sandage A., Freeman K., Stokes N., 1970, ApJ 160, 831

\bibitem[]{}
Sandage A., 1961, The Hubble Atlas of Galaxies, 
(Washington: Carnegie Institution)

\bibitem[]{}
Sandage A., Visvanathan N., 1978, ApJ 225, 742

\bibitem[]{}
Schweitzer F., Seitzer P., Faber et al., 1990, ApJ 364, L33

\bibitem[]{}
Smail I., Dressler A. Couch W. et al., 1997, ApJS 110, 213


\bibitem[]{}
van den Bergh S., 1976, ApJ 206, 883

\bibitem[]{}
van Dokkum P., Franx M., Kelson D., et al., 1998, ApJ, in press (astro-ph/9801190)

\bibitem[]{}
de Vaucouleurs G., 1959, Handbuch der Physik, vol. 53,
(Berlin: Springer Verlag)

\end{thebibliography}
\end{document}